\documentclass{rmaa}
\usepackage{amsmath}
\title{Interaction between the IGM and a dwarf galaxy}
\author{
V. Lora
\affil{Astronomisches Rechen-Institut, Zentrum f\"{u}r Astronomie der Universit\"{a}t Heidelberg, Germany}
A. C. Raga
  \affil{Instituto de Ciencias Nucleares, UNAM, M\'exico}
\and E. K. Grebel
\affil{Astronomisches Rechen-Institut, Zentrum f\"{u}r Astronomie der Universit\"{a}t Heidelberg, Germany}}

\fulladdresses{
\item Astronomisches Rechen-Institut, Zentrum f\"{u}r Astronomie der Universit\"{a}t Heidelberg,
  M\"{o}nchhofstr. 12-14, 69120 Heidelberg, Germany (vlora@ari.uni-heidelberg.de, grebel@ari.uni-heidelberg.de)
\item A. C. Raga: Instituto de Ciencias Nucleares, Universidad
Nacional Aut\'onoma de M\'exico, Ap. 70-543, 04510 D. F., M\'exico
(raga@nucleares.unam.mx)
}

\shortauthor{Lora, Raga \& Grebel}
\shorttitle{IGM interaction with a dwarf galaxy}

\SetVolume{40} \SetFirstPage{001} \SetYear{2014}
\ReceivedDate{\today} 
\AcceptedDate{Year Month Day} 

\resumen{ 
Las galaxias enanas son el tipo de objeto m\'as com\'un en el Universo y se cree
que contienen grandes cantidades de materia oscura.
Existen principalmente tres tipos morfol\'ogicos de galaxias enanas: el\'ipticas enanas,
esferoidales enanas, e irregulares enanas.
Las galaxias irregulares enanas son particularmente interesantes en la evoluci\'on de 
galaxias enanas, ya que los predecesores de las galaxias esferoidales enanas, pudieron 
haber sido muy similares a ellas. Entonces, un mecanismo ligado a la p\'erdida de gas
en las galaxias irregulares deber\'ia de ser observado, i. e. ram pressure stripping.
En este art\'\i culo estudiamos la interacci\'on entre el medio
interestelar y un medio intergal\'actico en movimiento relativo. Derivamos una
soluci\'on de plasm\'on de choque d\'ebil que corresponde al balance entre
la presi\'on post-choque a proa y la presi\'on estratificada del medio interestelar
(que suponemos sigue la estratificaci\'on de un halo de materia obscura gravitacionalmente
dominante). Comparamos nuestro modelo con simulaciones num\'ericas previamente
publicadas y con la nube de HI que rodea las galaxias irregulares enanas Ho~II y Pegaso. Mostramos
que este tipo de comparaci\'on provee una forma sencilla para estimar el n\'umero
de Mach del flujo.}

\abstract{
Dwarf Galaxies are the most common objects in the Universe and are believed to contain
large amounts of dark matter. 
There are mainly three morphologic types of dwarf galaxies:
dwarf ellipticals, dwarf spheroidals and dwarf irregulars. 
Dwarf irregular galaxies are particularly interesting in dwarf galaxy evolution, 
since dwarf spheroidal predecessors could have been very similar to them. Therefore,
a mechanism linked to gas-loss in dwarf irregulars should be observed, i.e. ram pressure stripping. 
In this paper, we study the interaction between the ISM of a dwarf 
galaxy, and a flowing IGM. We derive the weak-shock, plasmon solution corresponding to the
balance between the post-bow shock pressure and the pressure of the stratified ISM
(which we assume follows the fixed stratification of a gravitationally dominant dark
matter halo). We compare our model with previously published numerical simulations
and with the observed shape of the HI cloud around the Ho~II and Pegasus dwarf irregular 
galaxies. We show that such a comparison provides a straightforward way for estimating the Mach
number of the impinging flow.}


\begin{document}

\maketitle

\section{Introduction}

Dwarf galaxies are the most common objects in the Universe. These galaxies contain 
large amounts of dark matter (DM), and are better described by a cored DM mass
profile \citep{gilmore:07,governato:10,adams:14}. 
Dwarf galaxies appear in a variety of different morphological types \citep{grebel:01}.  
In galaxy groups, the most commonly occurring types are dwarf ellipticals (dE), dwarf 
spheroidal (dSph), and dwarf irregular (dIrr) galaxies.

Particularly, the DM content of dSphs may be as high as $90$\% 
(or more) of the total mass of the dwarf, even at the centre of the galaxy, which 
means that the dynamics in dSph galaxies are determined completely by the 
gravitational field of the DM halo \citep{binney:08} . 
dSphs have the peculiarity of containing almost no gas. If a large fraction of the gas 
content of a galaxy is removed, the star formation will decrease or even stop, and no 
stars would be formed with metallicities higher than the one of the gas at the moment 
of its removal. 

On the other hand, dIrrs are gas-rich, irregularly shaped galaxies with recent or ongoing 
star formation \citep{grebel:99}. They have low density and low surface brightness 
($\mu_{B} \approx 23$) and H~II regions that are superposed on an underlying diffuse 
structure of old stars with an exponential surface brightness profile 
(e. g. \citeauthor{pasetto:03}\citeyear{pasetto:03}).

% % %  TRANSICION DE dIrr A dSph  % % %
There are some dwarf galaxies in the Local Group (LG) which present intermediate features between
dIrr and dSph, suggesting that such dwarf galaxies could be in a transition 
state from dIrr to dSph \citep{grebel:03}. Transition dwarf galaxies (dTr) are 
dominated by old populations but contain gas and show recent star formation. They are 
situated at distances $>250$~kpc from the large spirals, and thus could be less influenced
by tidal stripping. 

A very good example of a transformation of a dwarf irregular into 
a dwarf elliptical galaxy by ram pressure stripping, is the dwarf galaxy IC3418. 
It presents a UV-bright tail comprised of knots, head–tail, and linear stellar 
features. \citeauthor{kenney:14} (\citeyear{kenney:14}) reported 
neither H$\alpha$ nor HI emission in the main body of the galaxy, but they detected 
$4\times10^{7}$~M$_{\odot}$ of HI from the IC3418 tail, which suggests that the HI 
in the main body of the galaxy is in the process of being stripped away. 

If indeed dIrrs evolve to become gas-free dSphs, there should be a mechanism or
mechanisms that removes the cold gas from dIrr galaxies. It has been suggested
that the gas loss could be controlled by the star formation rate and intense continuous 
galactic winds \citep{lanfranchi:07}, energy feedback from supernovae 
\citep{fragile:03,marcolini:06}, tidal stripping \citep{mayer:06}, and ram pressure 
stripping \citep{mori:00,marcolini:03,grebel:03,marcolini:06,grcevich:09}.

Ram pressure stripping has been studied numerically in great detail 
\citep{roediger:05,roediger:06a,roediger:06b,roediger:07} 
For example, \citeauthor{roediger:05} (\citeyear{roediger:05}) performed numerical 
simulations of disky galaxies in constant inter-cluster-medium winds, and found 
that even massive galaxies can lose a large amount of the disk gas, and even be 
completely stripped from their gas, if they are located in the cluster's centre. 
Moreover, in the subsonic regime, winds with small Mach numbers are more effective 
in getting rid of the gas, that the inclination does not play a major role for the mass 
loss \citep{roediger:06a}, that supersonic galaxies tend to develop  more irregular 
tails \citep{roediger:06b}, and that the tail density, length, and mass distribution,
of a disk galaxy orbiting a galaxy cluster, depends on the ram pressure as well as 
the galaxy's orbital velocity \citep{roediger:08}.

As a last example, \citeauthor{pasetto:03} (\citeyear{pasetto:03}) carried out N-body 
Tree-SPH simulations, in order to investigate whether a dIrr galaxy which tidally 
interacts with a MW-type galaxy may be reshaped into a dSph or an elliptical object. They 
conclude  that the transformation could be achieved, under some premises, on a time scale 
of $\gtrsim 4$~Gyr.

% % %  HOII Y PEGASUS  % % %
An example of a dIrr galaxy is the Ho~II dwarf.
Ho~II is a dIrr in the M81 group \citep{holmberg:50}.
Ho~II \citep{walter:07,oh:11} is very similar to the Small Magellanic Cloud (SMC) 
\citep{stanimirovic:99,stanimirovic:04} in absolute magnitude 
($M_{B}\sim-16.7$ and $-16.1$ respectively), HI content ($M_{HI}\sim6\times10^8$~M$_{\odot}$
and $\sim4\times10^8$~M$_{\odot}$ respectively) and total mass ($M=2.1\times10^{9}$~M$_{\odot}$
and $2.4\times10^{9}$~M$_{\odot}$).
The neutral gas centred on Ho~II shows a cometary morphology in its outer regions with
compressed contours perpendicular to its direction of motion through the IGM of the
M81 group, and extended, trailing tail structures on its opposite side \citep{bur02,bernard:12}.
It has to be noted that Ho~II is probably a companion of the outlying spiral NGC~2403 and its 
subgroup \citep{karachentsev:02}, which does not seem to participate in the ongoing interaction 
of the M81 subgroup. 

A second example is the dIrr galaxy Pegasus. Pegasus belongs to the LG of 
galaxies and it is located at a Galactocentric distance of $919$~kpc \citep{mcconnachie:05}. 
It has a total mass of $M\simeq3.3\times10^8$~M$_{\odot}$, and has a large 
amount of DM ($\sim70\%$) \citep{kniazev:09}. The Pegasus dIrr also presents the characteristic 
signature of ram pressure stripping \citep{young:03,mcconnechie:07}. 
Moreover, \citeauthor{mcconnechie:07} (\citeyear{mcconnechie:07}), conclude from the ram
pressure stripping in Pegasus, and its large distance from either the Milky 
Way or M31, that there is evidence for the existence of a LG intergalactic
medium.

% % %  COMO SE RELACIONA EL TRABAJO A HOII Y PEGASUS  % % %
In this paper we present a steady 
plasmon\footnote{The term plasmon was first introduced by 
\citeauthor{deyoung:66}(\citeyear{deyoung:66}), to refer to an emiting region 
roughly in hydrostatic equilibrium in a co-moving frame of reference.} model for the 
interaction of the ISM of a dwarf galaxy with a flowing IGM.
This model is of course appropriate for
galaxies with a dense enough ISM that can balance the ram pressure of the flowing
IGM. Two of the best examples of such a situation are the Ho~II and the Pegasus dwarf 
irregulars. In this paper we develop an analytical weak-shock plasmon model for
a dwarf galaxy in a IGM. We compare this analytical model with numerical simulations
and observations. We show that such a comparison provides a straightforward way for 
estimating the Mach number of the impinging IGM flow.

% % %  COMO SE DIVIDE EL ARTICULO  % % %
We derive a simple plasmon model along the lines of the model of 
\citet{deyoung:66} in Section~2, balancing the post-bow shock pressure with the pressure of
a stratified gas cloud. Our model has two important differences with respect to the plasmon of
\citet{deyoung:66}:
\begin{itemize}
\item the shock is not assumed to be strong,
\item the gas has a radial stratification which follows the profile of
a dark matter halo (which is assumed to be gravitationally dominant over the
stellar component of the dwarf galaxy).
\end{itemize}

Applications of the model to the case of a $1/R^2$ ISM pressure stratification
(Section~3) and to two distributions with flat cores (Section~4) are presented.
These solutions are then compared in Section~4 with the numerical simulations of ram-pressure
stripping flows of \citet{clo13} and with the observed shape of the HI clouds of Ho~II
\citep{bur02}, and Pegasus \citep{young:03}.

\section{The weak shock plasmon model}
\label{sec:1/r2}

The interaction between an initially spherically stratified ISM of a dwarf galaxy
and the intergalactic medium (through which the galaxy is travelling)
will result in a distortion of the ISM distribution, so that it is
progressively more flattened in the direction of the galaxy's
motion. This distortion progresses until a final
configuration is reached in which the post-bow shock pressure of the
intergalactic medium is approximately equal to the local pressure of
the ISM of the dwarf galaxy. Once this ``local pressure balance''
configuration is reached, the system will continue to evolve at a
slower rate through the entrainment of galactic ISM into the post-bow
shock sheath.

\begin	{figure*}
  \centering
  \includegraphics[width=0.6\textwidth]{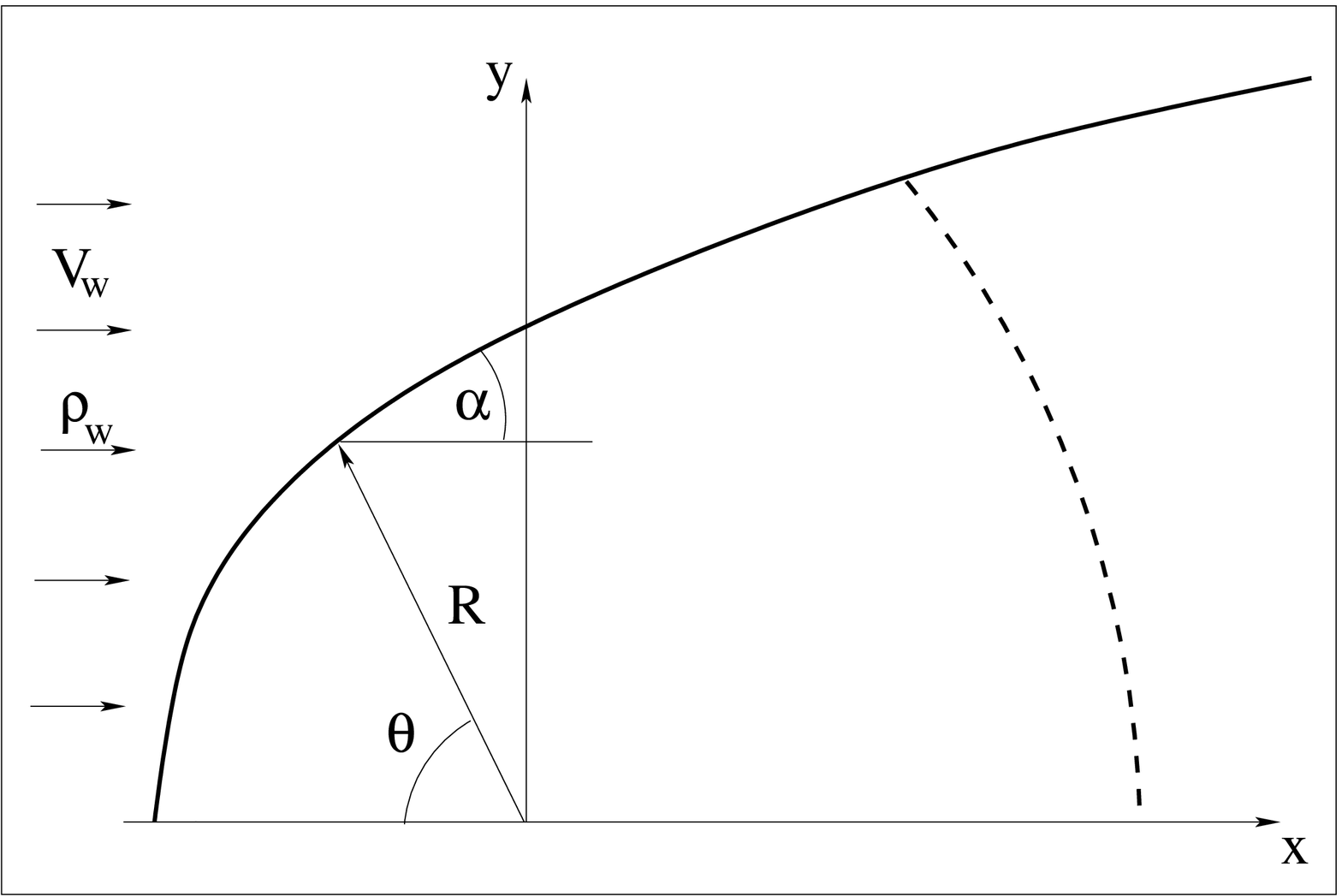}
  \caption{In this Figure, we show the IGM flow in a reference frame 
  moving with the dwarf galaxy, and with the $x$-axis pointing against 
  the direction of the motion of the galaxy. The intergalactic medium
  impinges on the dwarf galaxy with a uniform density $\rho_w$ and a
  velocity $v_w$, in the direction of positive $x$. The origin of the
  $(x,y)$ coordinate system coincides with the centre of the dwarf galaxy.
  The interaction region between
  IGM and the ISM of the dwarf galaxy is
  assumed to be thin, and is represented by the thick black curve. 
  The $\theta$ angle is measured from the $-x$ direction, and the $\alpha$
  angle is the angle between the locus of the interaction region and the 
  symmetry axis.}
  \label{fig:FIG1}
\end{figure*}

Following \citet{deyoung:66}, who studied the case of a
decelerating gas clump (see Appendix 1), we derive a ``plasmon
solution'', assuming that the region of the extragalactic/galactic
medium interaction is narrow, and neglecting the centrifugal pressure
of the material flowing along the interaction sheath (see \citeauthor{canto:98}
\citeyear{canto:98}). The structure of the flow is shown in the schematic diagram
of Figure~\ref{fig:FIG1}.

This diagram shows the flow in a reference frame moving with the dwarf
galaxy, and with the $x$-axis pointing against the direction of the motion
of the galaxy. In this reference frame, the intergalactic medium
impinges on the dwarf galaxy (with a uniform density $\rho_w$ and a
velocity $v_w$) in the direction of positive $x$. The origin of the
$(x,y)$ coordinate system coincides with the centre of the galaxy, so
that the unperturbed spherical pressure stratification of the galactic ISM is
$P(R)$, where $R$ is the spherical radius. The interaction region is
assumed to be thin, and is represented by the thick curve (which
corresponds to either the bow shock or to the outer boundary of the
unperturbed galactic ISM). In the derivation of the model we use the
angles $\theta$ (measured from the $-x$ direction) and $\alpha$ (the
angle between the locus of the interaction region and the symmetry axis).

For a general (i.e., not necessarily strong) shock, the post-bow shock
pressure is given by:
\begin{equation}
\label{pps}
P_{PS} = \frac{2}{ \gamma+1} \rho_{w} v_{w}^{2} \sin^{2}(\alpha) - \frac{\gamma -1}{\gamma +1} P_{w} \mbox{ ,}
\end{equation}
where $\rho_w$, $v_w$ and $P_w$ are the density, velocity and pressure (respectively)
of the impinging intergalactic medium, $\gamma$ is the specific heat
ratio and the angle $\alpha$ is shown
in Figure~\ref{fig:FIG1}. It is clear that in the weak shock limit (i.e., $v_w\to
c_w=\sqrt{\gamma P_w/\rho_w}$) we correctly obtain $P_{PS}=P_w$.

Equation (\ref{pps}) is a straightforward generalization to the case
of a general shock jump (strong or weak)
of the relation used by De Young \& Axford (1967) and Dyson (1975) to
model strong (i.e., hypersonic) bow shocks.

Now, if $y(x)$ is the shape of the interaction region (see Figure~\ref{fig:FIG1}),
we have the relation:
\begin{equation}
\label{dxy1}
\sin^2\alpha=\left[1+\left(\frac{dx}{dy}\right)^2\right]^{-1}\,.
\end{equation}
Also, the $dx/dy$ derivative is related to the interaction surface in
spherical coordinates $R(\theta)$ through:
\begin{equation}
\label{dxy2}
\frac{dx}{dy}=\frac{\frac{-dR}{d\theta}+ R \tan \theta}{\tan \theta \frac{dR}{d\theta}+R} \mbox{ .}
\end{equation}
Combining equations (\ref{dxy1}-\ref{dxy2}) we then obtain:
 \begin{equation}
 \label{eq:alpha}
 \sin^{2}\alpha = \frac{\left(\tan \theta \frac{dR}{d\theta} + R\right)^2} {[( \frac{dR}{d\theta}  )^2 + R^2   ] } \mbox{ .}
 \end{equation}

Finally, the $P_{PS}=P(R)$ condition of balance between the post-shock pressure
and the spherically stratified pressure of the galactic ISM is:
\begin{equation}
\label{pr}
P(R)= \frac{2}{\gamma +1} \rho_{w} v_{w}^2 \sin^2\alpha - P_{w} \frac{\gamma -1 }{\gamma +1}  \mbox{ ,}
\end{equation}
where we have used Equation (\ref{pps}). We then solve this equation
for $\sin^2\alpha$, and combine it with Equation (\ref{pr}) to obtain:
  \begin{equation}
 \label{quad}
  \left( \tan \theta \frac{dR}{d\theta} + R\right)^2 \cos^2\theta = F(R) \left[\left(\frac{dR}{d\theta}\right)^2+R^2 \right] \mbox{ ,}
  \end{equation}
where we have defined
 \begin{equation}
  \label{eq:alpha2}
 F(R) \equiv [  (\gamma +1 ) P(R) + P_{w} (\gamma +1) ] \frac{1}{ 2 \rho_{w} v_{w}^2}   \mbox{ .}
 \end{equation} 

Equation (\ref{quad}) is a quadratic equation for $dR/d\theta$, which
can straightforwardly be solved to obtain:
 \begin{equation}
\label{plasmon}
 \frac{dR}{d\theta} = \frac{R}{F(R) - \sin^2\theta}  \left\{     \frac{\sin 2\theta}{2} - \sqrt{F(R) [ 1-F(R) ] }    \right\}  \mbox{ .}
 \end{equation}
In this way, we obtain the differential equation that describes the
$R(\theta)$ shape (see Figure~\ref{fig:FIG1}) of the interaction region
between the stratified galactic ISM and the impinging intergalactic 
medium. We should note that the negative branch of the solution has 
been chosen (because this branch gives the physical solution).

One then has to integrate Equation (\ref{plasmon}) with the boundary
condition $R(\theta=0)=R_0$ (the stagnation region radius), and
requiring the shock to be perpendicular to the $x$-axis in the
stagnation point. From Equation (\ref{pr}), we see that this condition
can be written as:
 \begin{equation}
 P(R_{0})=\frac{2}{\gamma +1 } \rho_{w}  v_{w} - \frac{\gamma - 1}{\gamma +1} P_{w} \mbox{ ,}
 \end{equation}
 and substituting this pressure in equation (\ref{eq:alpha2}) we obtain:  
  \begin{equation}
  F(R_{0})=\frac{1}{2}  [  2 -  (\gamma -1) \frac{P_{w}}{\rho_{w}
    v_{w}^2} + \frac{(\gamma -1)}{\gamma M_{w}^2} ] = 1 \mbox{ ,}
\label{fr0}
  \end{equation}
where $M_{w}=v_w/\sqrt{\gamma P_w/\rho_w}$ is the Mach number of the
impinging intergalactic medium.

It is clear that Equation (\ref{plasmon}) only has to be integrated
until the value of $\theta_m$ for which $\sin \alpha=1/M_w$ (see 
Figure~\ref{fig:FIG1}). At this point, the shock becomes a sound wave, and detaches from
the surface of the plasmon. This can be seen from the fact that if we have smaller
values of $\alpha$, the post-shock pressure has values $P_{PS}<P_w$
(see Equation \ref{pps}), so that the shock has been changed into an
unphysical ``expansion jump''. Therefore, for $\theta>\theta_m$ the
shock becomes a detached conical surface (with half-opening angle
$\alpha_m=\sin^{-1}(1/M_w)$), and the surface of the plasmon is the
circular surface determined from the $P(R)=P_w$ condition. This
circular surface is shown with a dashed line in the schematic diagram
of Figure~\ref{fig:FIG1}.
 
 % % % % % % % % % % % % % % % % % % % % % % % % % % % % % % % % % % % % % % % % % % % % % % % % % % % % % % % % % % % % % % % % % % % % % % % % %
 \section{The $P(R)\propto R^{-2}$ case}
 \label{sec:r2}

Let us now consider the case of a galaxy with an ISM pressure
stratification of the form
\begin{equation}
P(R)=\frac{A}{R^2}\,,
\label{pr2}
\end{equation}
where $A$ is a constant.

In terms of a dimensionless radius $r=R/R_0$ (where $R_0$ is the stagnation
region radius), Equation (\ref{plasmon}) becomes:
 \begin{equation}
\label{aplasmon}
 \frac{dr}{d\theta} = \frac{r}{F(r) - \sin^2\theta}  \left\{     \frac{\sin 2\theta}{2} - \sqrt{F(r) [ 1-F(r) ] }    \right\}  \mbox{ ,}
 \end{equation}
with
\begin{equation}
F(r)=\frac{1}{2}  \left\{ \left[ 2- \frac{(\gamma -1)}{\gamma}
    \frac{1}{M_{w}^2}  \right] \frac{1}{r^2} +
\frac{(\gamma-1)}{\gamma} \frac{1}{M_{w}^2}  \right\} \mbox{ .}
\label{f2}
\end{equation}
as obtained from Equations (\ref{eq:alpha2}), (\ref{fr0}) and
(\ref{pr2}). As could be expected, we have not found an analytic
integral for the differential equation obtained combining equations (\ref{aplasmon}-\ref{f2}).

\begin{figure*}
  \centering
  \includegraphics[width=0.75\textwidth]{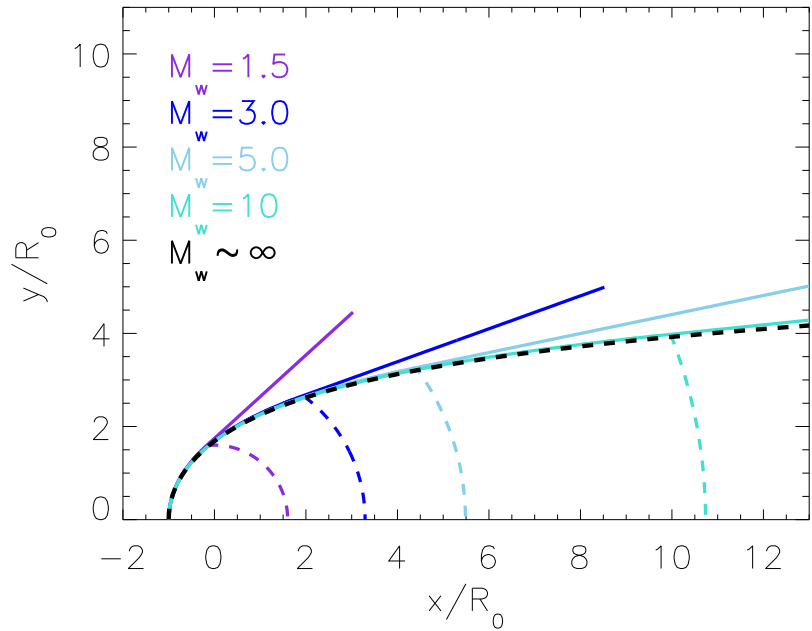}
  \caption{In this Figure we show the plasmon solutions for the values
  of the Mach number $M_w=1.5, 3, 5, 10$ and $\infty$, of the impinging 
  intergalactic medium (considering $\gamma=5/3$).
  In this case, the dwarf galaxy has an ISM pressure stratification of 
  the form $P(R)\propto R^{-2}$. The $x$-axis is pointing against the 
  direction of the motion of the galaxy, and $R_{0}$ is the stagnation 
  region radius.}
  \label{fig:FIG2}
\end{figure*}

We therefore integrate numerically Equation (\ref{aplasmon}) for different
values of the Mach number $M_w$ of the impinging flow. The integration
is carried out until the surface of the plasmon reaches the Mach angle
$\alpha_m=\sin^{-1}(1/M_w)$ (at $\theta=\theta_m$), and for larger values of $\theta$ the
shock is prolonged with the sonic slope and the plasmon is closed with
a spherical surface (as described in the last paragraph of Section~2).
The results of this exercise are shown in Figure~\ref{fig:FIG2}.

In this Figure, we show the plasmon solutions obtained for different values
of the Mach number $M_w$ of the impinging intergalactic medium (and
for an ideal monoatomic gas, i.e. $\gamma=5/3$). In the head of the plasmons, 
the shock wave coincides 
with the surface of the plasmon, but it detaches when the normal component 
of the pre-shock velocity becomes sonic. Downstream of this point, the 
surface of the plasmon takes a constant pressure, spherical shape, and the 
shock wave is a straight, sonic wave. The region of the plasmon upstream of the
``sonic point'' (at $\theta=\theta_m$, see above) is very similar to
the solutions obtained for higher Mach number plasmons.
Therefore, a good approximation to the shape of the plasmon can be
obtained by taking the $M_w\to \infty$ solution of Equations
(\ref{aplasmon}-\ref{f2}) and cutting it off at the value of $\theta_m$
corresponding to the actual value of $M_w$ of the impinging
intergalactic medium.

We have been unable to obtain an analytic solution of Equations
(\ref{aplasmon}-\ref{f2}), even in the simpler $M_w\to \infty$
case. However, it is possible to derive an approximate analytic form
in the following way.

The problem that we are considering is similar
to the problem of the ram pressure balance between a spherical,
constant velocity wind and a uniform, streaming environment. 
\citet{dys75} showed that this problem has the analytic solution
$r_{Dy}=\theta/\sin \theta$ (where $r_{Dy}$ is the spherical radius in
units of the stagnation radius). We therefore propose an approximate
analytic solution to our plasmon solution of the form $r_{Dy}^\beta$,
and determine $\beta$ from a fit to the $M_w\to \infty$ numerical
solution of Equations (\ref{aplasmon}-\ref{f2}). In this way we obtain
the approximate analytic solution:
\begin{equation}
r_a=\left(\frac{\theta}{\sin \theta}\right)^{1.17}
\label{ra}
\end{equation}
for the $M_w\to \infty$ case (which is independent of $\gamma$).

\begin{figure*}  
  \centering
 \includegraphics[width=0.6\textwidth]{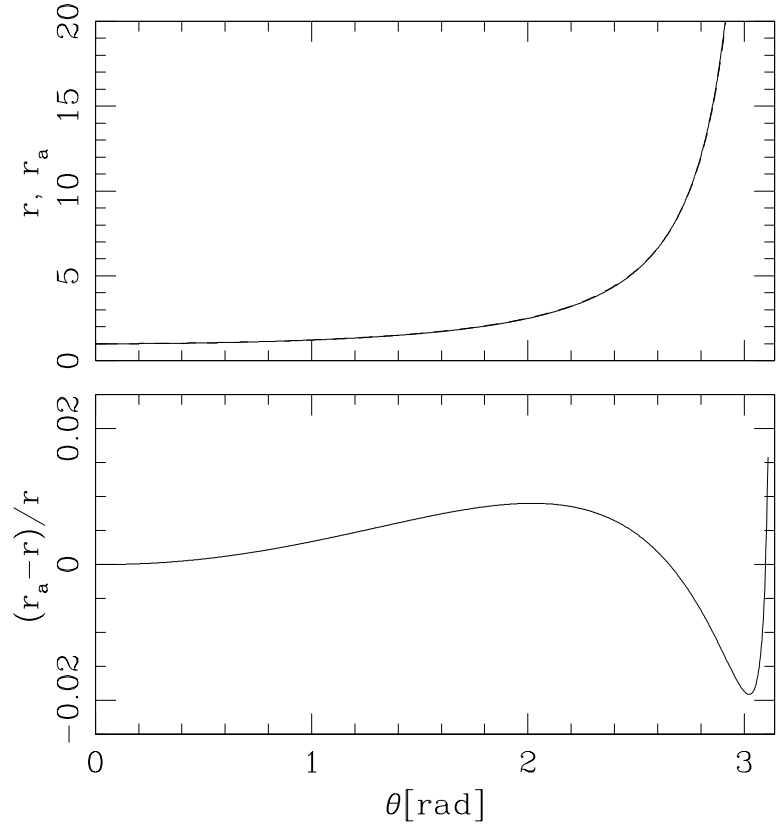}
  \caption{In this Figure, we show a comparison between the $r_a(\theta)$ approximate
  analytic solution (see Equation~\ref{ra}) and the $M_w\to\infty$ ``exact'' numerical 
  solution (upper panel), together with the associated relative error (bottom panel)
  as a function of $\theta$ (see Figure~1).}
  \label{fig:FIG3}
\end{figure*}

In Figure~\ref{fig:FIG3}, we show a comparison between $r_a(\theta)$ and the $M_w\to
\infty$ numerical solution, together with the associated relative
error. We see that the approximate analytic solution reproduces the
numerical (``exact'') solution to within $\sim 2$\%\ within the whole
$\theta=0\to \pi$ range. It is then possible to obtain approximate
solutions for all $M_w$ by truncating the approximate analytic
solution (Equation \ref{ra}) at the point in which the slope becomes
sonic (with $\sin \alpha_m=1/M_w$, see above).

\begin{figure*}  
  \centering
 \includegraphics[width=0.7\textwidth]{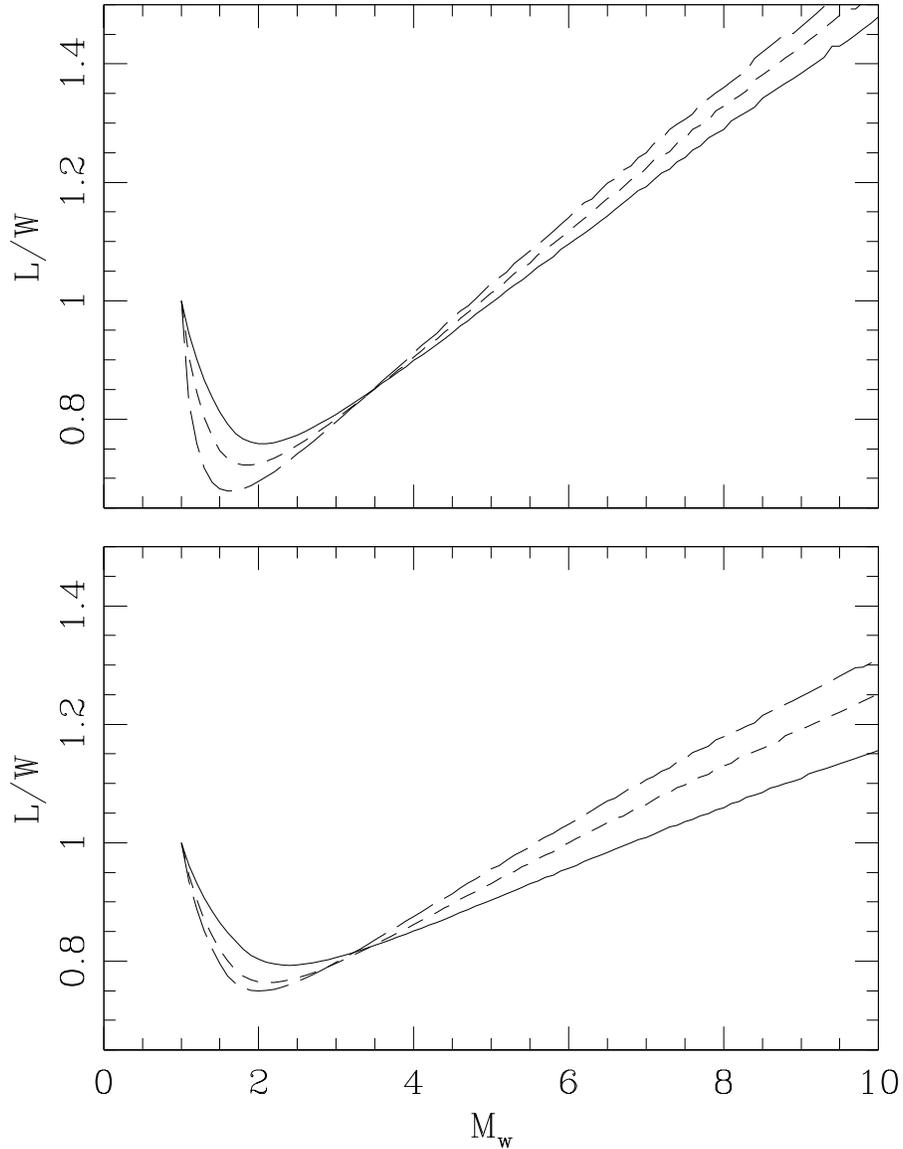}
  \caption{In this Figure we show the results of the numerical solutions
  of Equations (\ref{aplasmon}-\ref{f2}). We compute the length-to-width 
  ratio $L/W$ as a function of the Mach number $M_w$ of the flowing IGM.
  We measure $L$ along the symmetry axis, from the stagnation point to the point 
  in which the spherical back side of the plasmon intersects the axis. 
  The width $W$, is twice the maximum cylindrical radius attained by the plasmon
  solution. Top frame: the results obtained for the core-less pressure stratification
  (Equation \ref{prc}) are shown
  with the solid line, and the short-dash and long-dash lines correspond to
  stratifications with core radii of $R_{core}=R_0$ and $2R_0$ (respectively, see Equation \ref{prc}). 
  Bottom frame: $L/W$ as a function of Mach number for plasmons with a NFW 
  pressure profile (see Equation \ref {nfw}) with $R_s=0$ (solid line)	,
  $R_0$ (short dash) and $2R_0$ (long dash line).}
  \label{fig:FIG4}
\end{figure*}

Now, from the numerical solutions of Equations (\ref{aplasmon}-\ref{f2})
we compute the length-to-width ratio $L/W$ (where $L$ is measured
along the symmetry axis from the stagnation point to the point in
which the spherical back side of the plasmon intersects the axis, and
the width $W$ is twice the maximum cylindrical radius attained by the
plasmon solution) as a function of the $M_w$ Mach number (which is the
only free parameter of the dimensionless plasmon solution). The
results of this exercise are shown in Figure~\ref{fig:FIG4} (very similar results
being obtained from the approximate analytic solution described
above).

We see that the length-to width ratio $L/W$ has a minimum at
$M_w\approx 2$, increasing to a value of 1 (at $M_w=1$) for smaller
values of $M_w$, and increasing monotonically for $M_w>2$ 
(see Figure~\ref{fig:FIG4}). Actually, for $M_w<2$ our plasmon model is probably not
applicable, since for Mach numbers approaching unity the stand-off
distance of the shock in the stagnation region becomes comparable to
the radius of the plasmon, so that the ``thin interaction region''
approximation of our model is not valid. On the other hand, for
$M_w>2$ our plasmon model should be a reasonable approximation to the
real flow.

Therefore, for a ram-pressure confined plasmon in which the impinging
intergalactic medium has a Mach number $M_w>2$, the curve shown in
Figure~\ref{fig:FIG2} can be used to determine $M_w$ from the
length-to-width ratio of the plasmon. Of course, due to
projection effects the observed
length-to-width ratio is only a lower boundary of the intrinsic $L/W$
of the plasmon, and therefore the observations only determine a lower
boundary for the possible value of the Mach number $M_w$ of the flow.

% % % % % % % % % % % % % % % % % % % % % % % % % % % % % % % % % % % % % % % % % % % % % % % % % % % % % % % % % % % % % % % % % % % % % % %
\section{Distributions with cores}
\label{sec:core}

Since the dark matter component in dwarf irregular galaxies is dominant, 
we suppose that the gas within the dwarf galaxy follows the gravitational 
potential of the dark matter, therefore having the same density mass 
distribution. The $P(R)\propto R^{-2}$ explored in section 2 is a singular profile, but
dark matter haloes of dwarf galaxies are more commonly modeled either with a modified
profile with a core radius (approximating the non-singular isothermal
sphere solution, see, e.g., Shapiro et al. 1999)
and the NFW profile (see Navarro, Frenk \& White 1996).

We first explore the case of a dwarf galaxy with an ISM 
pressure stratification of the form:
\begin{equation}
P(R)=\frac{A}{(R^2+R_{core}^2)}\,,
\label{prc}
\end{equation}
with a core radius $R_{core}$, which is the simplest analytic approximation
to the non-singular isothermal sphere (see, e.g., Hunter 2001).

Substituting this pressure
stratification in Equation (\ref{eq:alpha2}) we obtain
  \begin{equation}
  \label{eq:F_core}
 F(R)=\frac{1}{2} \left[  \frac{(\gamma+1)A}{\rho_{w} v_{w}^2}
 \frac{1}{R^2+R_{core}^2} +
\frac{(\gamma-1)}{\gamma} \frac{1}{M_{w}^2}  \right]\mbox{ .}
 \end{equation}
Using the $F(R_0)=1$ boundary condition (see Equation \ref{fr0}), we
then have:
%\begin{equation}
\begin{multline}
F(r) = \frac{1}{2} \Biggl[  \Biggl(  2 - \frac{(\gamma-1)}{\gamma}   \frac{1}{M_{w}^2}  \Biggr)
\Biggl(\frac{1+r_c^2}{r^2+r_{0,c}^2}\Biggr)  \\
+ \frac{(\gamma-1)}{\gamma} \frac{1}{M_{w}^2}\Biggr] \,,
\label{frc}
\end{multline}
%\end{equation}

with $r=R/R_0$ and $r_c=R_{core}/R_0$, where $R_0$ is the stagnation region radius of the plasmon.

>From numerical integrations of Equations
(\ref{aplasmon}) and (\ref{frc}), we compute the length to width ratio $L/W$
as a function of the Mach number $M_w$ of the flow for models with core
radii $R_{core}=R_0$ and $R_{core}=2R_0$. From Figure~4, we see that
the predicted values of $L/W$ only have relatively small deviations
from the values obtained for the core-less pressure distribution (see
section 3).

We also compute models for a NFW mass density profile \citep{nav96}
given by

\begin{equation}
\rho(R)=\frac{\rho_{0}}{\frac{r}{R_s}(1+\frac{r}{R_s})^2} \mbox{ .}
\label{rho_nfw}
\end{equation}

Thus, the pressure of the NFW profile can be writen as:
\begin{equation}
P(R)=\frac{A}{R(R+R_{s})^2} \mbox{ .}
\label{nfw}
\end{equation}

We compute the NFW pressure for a $R_{s}=0$, $R_0$ and $2R_0$. 
The resulting $L/W$
vs. $M_w$ relations are shown in the bottom panel of Figure~4. From
this Figure we see that the length-to-width ratios only differ
substantially from the $L/W$ values obtained for the $1/R^2$ pressure
law for relatively high ($M_w>4$) Mach numbers. 

>From the results of this section, we see that for the three studied
ISM pressure distributions (Equations \ref{pr2}, \ref{prc} and
\ref{nfw}) we obtain a minimum value $L/W\approx 0.7\to 0.8$ for the
predicted length-to-width ratio of the plasmon shape, and that this
minimum occurs for a Mach number $M_w\approx 1.7\to 2.3$ of the impinging flow.

\section{Comparison with simulations and observations}

It is possible to compare our plasmon models with the simulations of
\citet{clo13}. These authors computed 3D numerical simulations of
ram-pressure stripping of a structure with a NFW
profile (Equation \ref{nfw}) for an impinging flow with Mach numbers
$M_w=0.9$, 1.1 and 1.9.

The simulations
of \citet{clo13} produce a recognizable core and an extended, lower
density wake of stripped material (see their Figure~2). The cores of
the $M_w=0.9$ and 1.1 simulations have length to width radii $L/W\sim
1$, which is consistent with the $M_w\to 1$ limit of our plasmon model
(see Figure~4), though our model is not clearly applicable to such low
Mach numbers. The $M_w=1.9$ simulation of \citet{clo13} produces a
flattened core, with $L/W\sim 0.7$ (see the $t=2512$ Myr time frame
of their Figure~3). This length-to-width ratio is clearly consistent
with the $L/W$ values predicted from our plasmon model for $M_w=2$
(see the bottom frame of Figure~4).

Let us now compare our plasmon model with the HI maps of Ho~II.
>From the map of \citet{bur02}, using the $N_H=6\times
10^{19}$cm$^{-2}$ contour (see their Figure~3), we calculate a
length-to-width ratio $L/W=0.8\pm 0.1$. This value is consistent with
the predictions from our plasmon model for any of the pressure
distributions that we have studied, for Mach numbers $M_w\sim 1.5\to
3$ (see the two frames of Figure~4).

This comparison between the HI emission of Ho~II and our model is
strictly correct only if the motion of Ho~II lies on the plane
of the sky. In order to evaluate the possible projection effects, we
take a $M_w=2$, core-less pressure stratification (Equation \ref{pr2})
plasmon model, and rotate its axis at an arbitrary angle $\phi$ with respect to
the plane of the sky.

The projected length-to-width ratios computed from the
rotated plasmon model as a function of $\phi$ are shown in Figure~5. 
>From this Figure, we see that the projected $L/W$ value slowly
increases with increasing $\phi$ up to an angle $\phi\sim 40^\circ$,
and then rises more rapidly to the expected $L/W=1$ value for
$\phi=90^\circ$ (that is, with the symmetry axis along the line of
sight). From this exercise we conclude that the straightforward
comparison of the $L/W$ ratio of Ho~II with the (unprojected) values
predicted from our plasmon models is probably reasonable, since
projection effects become important only
for relatively high values of $\phi$. Such high values of the orientation
angle have low probabilities of occurrence, since they correspond to
directions pointing into a small solid angle.

We now compare the plasmon model with the Pegasus dwarf HI maps of
\citeauthor{young:03} (\citeyear{young:03}) (see their Figure~5). 
We take the contours $N_H=4\times 10^{19}$cm$^{-2}$ and 
$N_H=1.6\times10^{20}$cm$^{-2}$, and computed a very similar 
length-to-width ratio for both contours, $L/W=1.5\pm 0.1$ and $L/W=1.4\pm 0.1$ 
respectively.
 
If we compare this value of $L/W$ of Pegasus dwarf, with the predictions 
from our plasmon model (top panel of Figure~4), we deduce a high Mach 
number of $M_w\approx 9$ for the LG IGM.\\ \\

% % % % % % % % % % % % % % % % % % % % % % % % % % % % % % % % % % % % % % % % % % % % % % % % % % % % % % % % % % % % % % % % % % % % % % % % % %
\section{Conclusions}
\label{sec:conclusions}

We describe a simple plasmon model for the interaction of a dwarf galaxy
with a stratified ISM (following the stratification of a dark matter halo) and
a surrounding ISM flowing at a relatively low Mach number. We use this model in order
to compute the resulting confined shapes for different forms of the
ISM pressure stratification.

>From these models we have calculated the length-to-width ratios of the plasmon shapes,
finding that in all cases a minimum $L/W$ value of $\sim 0.8$ is obtained for
a $M_w\sim 2$ Mach number for the flowing IGM. We find that the predicted values of
$L/W$ are consistent with the aspect ratios of the central cores of the 3D numerical
simulations of \citet{clo13}. In addition, the fact that numerical simulations do produce an
approximately steady, plasmon-like core (unlike the \citeauthor{deyoung:66}'s \citeyear{deyoung:66}
 ``decelerating cloud plasmon'' which is not reproduced in numerical simulations) indicates that our
plasmon solution is indeed applicable to real astrophysical flows.

Finally, if we look at the shape of the HI cloud surrounding the Ho~II dwarf spheroidal,
we find that it has a length-to-width ratio $\sim 0.8$, consistent with the
predicted plasmon shapes for $M_w\sim 2$. We argue that the observed shape of the Ho~II cloud
is unlikely to be strongly affected by projection effects, so that this is a valid comparison
with our model. 

For the Pegasus dIrr, we find a length-to-width ratio  $\sim 1.5$ which indicates
a very high Mach number ($M_{w}\sim9$) for the intra-cluster medium. The latter 
high value of the Mach number seems to be unlikely for a galaxy at the location
of Pegasus. On the other hand, the LG intergalactic medium might be clumpy,
therefore it is possible that Pegasus is moving through a higher density and
lower temperature region, compared to other dwarf galaxies in the LG \citep{mcconnechie:07}, 
and as a consequence would present a higher Mach number.

Clearly, these comparisons between the plasmon and specific
  objects is correct only if the velocity of the galaxies with respect
  to the intergalactic medium lies close to the plane of the sky. In
  Section 5, we argue that projection effects should be small (so that
  a direct comparison with the models without projection corrections
  is appropriate) for angles between the direction of motion and the
  plane of the sky $\phi<40^\circ$. Higher values of $\phi$ have lower
probabilities of occurrence (orientation angles $\phi>40^\circ$ have probabilities
of occurrence $p=1-\sin 40^\circ=0.36$), but of course could be
relevant for specific objects. An important projection effect
(obtained for $\phi>40^\circ$) would result in an underestimate of the
Mach number of the flow.

We should note that an application of our plasmon model to observed galaxies is
clearly only relevant if the ram-pressure stripping mechanism is
actually active in the observed objects. As discussed in Section~1, this is 
only one of the possible mechanisms for stripping the ISM
from dwarf galaxies.

Our present semi-analytic model gives an approximate prediction of the
shape expected for the ISM of a dwarf galaxy subject to the
gravitational force of a dark matter potential and the ram pressure of
an impinging intergalactic medium. This model can be used to estimate
the flow parameters (in particular, the Mach number of the impinging
flow) that would produce the observed shape. These results could then
be fed into a numerical simulation (e.g., like the ones of
Close et al. 2013) in order to obtain more detailed predictions of the
observational characteristics of the flow (e.g., 21cm line
emission maps and velocity channel maps). A comparison between such
predictions and observations would be useful for showing whether or
not the ram-pressure stripping mechanism is actually responsible for
the observed structure of the ISM.

\section*{Acknowledgements}
\acknowledgments
V.L. gratefully acknowledges support from the FRONTIER grant
from the University of Heidelberg.
AR acknowledges support from the CONACyT grants
101356, 101975 and 167611, and the DGAPA-UNAM grants
IN105312 and IG100214. We also acknowledge an anonymous referee for
helpful comments.

%*************************************************
%           ----   FIGURAS   ----                %
%*************************************************
%
\begin{figure*}
  \centering
  \includegraphics[width=0.6\textwidth]{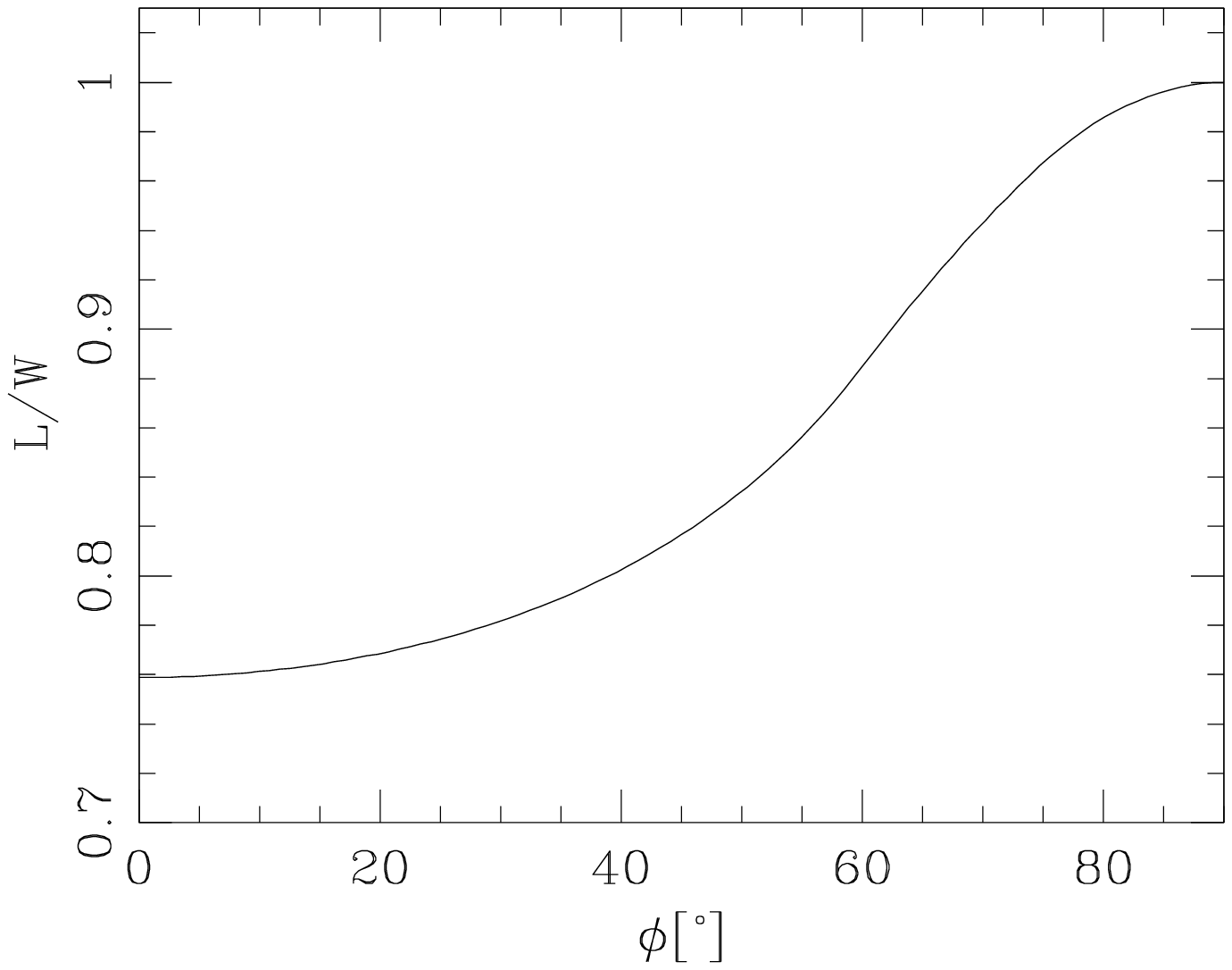}
  \caption{Predicted $L/W$ ratio for a core-less, inverse square pressure
  stratification (Equation \ref{pr2}), $M_w=2$ plasmon as a function of the angle $\phi$
  between the symmetry axis and the plane of the sky.}
  \label{fig:FIG5}
\end{figure*}
%
%%%%%%%%%%%%%%%%%%%%%%%%%%%%%%%%%%%%%%%%%%%%%%%%%%%%%%%%%%%%%%%%%%%%%%%
 \appendix 
 \section[]{The plasmon of De Young \&  Axford}
\citet{deyoung:66} derived the pressure balance plasmon shape
for a decelerating gas clump. In this problem, if the plasmon is
isothermal and has a constant deceleration $a$, its internal pressure has
an exponential stratification:
\begin{equation}
P(x)=P_{0} e^{-x/H} \mbox{ ,}
\end{equation}
where $x$ is the distance along the symmetry axis measured from the
stagnation point, $H=RT/\mu a$ (with $T$ the temperature, $R$ the gas constant and
$\mu$ the mean molecular weight) is the pressure scale-height and
$P_0$ is the pressure at the stagnation point. The shape of the
plasmon is then derived by setting $P(x)=P_{PS}$ (the post-shock
pressure). Using Equations (\ref{pps}) and (\ref{dxy1}), this
condition gives:
\begin{equation}
\frac{dx}{dr} = \sqrt{\frac{1}{\kappa + [1-\kappa]e^{-x/H}} -1}  \mbox{ ,}
\label{da}
\end{equation}
where $\kappa=1/m_w^2$ with
\begin{equation}
m_w=\sqrt{\frac{2\gamma}{\gamma-1}}M_w\,,
\label{mw}
\end{equation}
where $M_w$ is the Mach number of the streaming environment. The
plasmon solution of \citet{deyoung:66} is derived from Equation
(\ref{da}) with $\kappa=0$. For non-zero $\kappa$ his equation has the analytic solution:
$$\frac{r}{H} = \tan^{-1} \left(  \frac{2 \sqrt{1-u} \sqrt{a+u}}{2u+a-1}  \right) + $$
\begin{equation}
\sqrt{a}\, \log\left[\frac{2a+u(1-a)+2\sqrt{a} \sqrt{1-u}
    \sqrt{a+u}}{au}\right]\mbox{ ,}
\label{da2}
\end{equation}
where
\begin{equation}
a=\frac{1}{m_{w}^{2}-1}=\frac{\kappa}{1-\kappa}\,.
\end{equation}
It is straightforward to show that for $m_w\to \infty$ Equation
(\ref{da2}) coincides with the plasmon solution of \citet{deyoung:66}.

% 
% Spectroscopic observations of bright quasars show that the mean
% number density of Ly$\alpha$ forest lines, which satisfy certain
% criteria, evolves like $\rmn{d}N/\rmn{d}z=A(1+z)^\gamma$, where
% $A$ and~$\gamma$ are two constants.  Given the above intrinsic
% line distribution we examine the probability of finding large gaps
% in the Ly$\alpha$ forests.  We concentrate here only on the
% statistics and neglect all observational complications such as the
% line blending.

\end{document}